\documentclass[
	11pt, 
	a4paper, 
	]{article}

\usepackage{mathtools}
\usepackage{bbm}
\usepackage{tikz}
\usepackage{amssymb} 
\usepackage{amsmath} 
\usepackage{geometry} 
\usepackage{comment}
\usepackage{setspace} 
\usepackage{natbib} 
\usepackage{hyperref} 
\usepackage{ntheorem} 
\usepackage{floatrow}
\newfloatcommand{capbtabbox}{table}[][\FBwidth]

\theoremstyle{break} 

\newtheorem{theorem}{Theorem}
\newtheorem{corollary}{Corollary}

\newtheorem{proposition}{Proposition}

\newenvironment{proof}[1][Proof]{\noindent\textbf{#1. }}{\ \rule{0.5em}{0.5em}}

\begin{document}

\begin{titlepage}
\title{
	Vote Delegation and Misbehavior
}

\author{
	Hans Gersbach\\
	\normalsize CER-ETH and CEPR\\
	\normalsize Z\"{u}richbergstrasse 18\\
	\normalsize 8092 Zurich, Switzerland\\ 
	\normalsize \href{mailto:hgersbach@ethz.ch}{hgersbach@ethz.ch}
	\and
	Akaki Mamageishvili\\
	\normalsize CER-ETH\\
	\normalsize Z\"{u}richbergstrasse 18\\
	\normalsize 8092 Zurich, Switzerland\\ 
	\normalsize \href{mailto:amamageishvili@ethz.ch}{amamageishvili@ethz.ch}
	\and
	Manvir Schneider\\
	\normalsize CER-ETH\\
	\normalsize Z\"{u}richbergstrasse 18\\
	\normalsize 8092 Zurich, Switzerland\\ 
	\normalsize \href{mailto:manvirschneider@ethz.ch}{manvirschneider@ethz.ch}
	}

\date{Last updated: \today}

\maketitle

\begin{abstract}
    We study vote delegation with ``well-behaving" and ``misbehaving" agents and compare it with conventional voting. Typical examples for vote delegation are validation or governance tasks on blockchains. There is a majority of well-behaving agents, but they may abstain or delegate their vote to other agents since voting is costly. Misbehaving agents always vote. We compare conventional voting allowing for abstention with vote delegation. Preferences of voters are private information and a positive outcome is achieved if well-behaving agents win. We illustrate that vote delegation leads to quite different outcomes than conventional voting with abstention. In particular, we obtain three insights: First, if the number of misbehaving voters, denoted by $f$, is high, both voting methods fail to deliver a positive outcome. Second, if $f$ takes an  intermediate value, conventional voting delivers a positive outcome, while vote delegation fails with probability one. Third, if $f$ is low, delegation delivers a positive outcome with higher probability than  conventional voting. Finally, our results characterize worst-case outcomes that can happen in a liquid democracy.
	\bigskip
\end{abstract}
\thispagestyle{empty}
\end{titlepage}

\pagebreak \newpage

\setcounter{page}{2}
\linespread{1.6}

\section{Introduction}\label{sec:introduction}

We study vote delegation when preferences are private information and voting is costly. Vote delegation plays a crucial role in two contexts---blockchain governance and so-called ``liquid democracy", that is, a democracy in which the electorate can choose between voting itself or delegating its right to vote before each collective decision.

Let us look at the standard problem of blockchains with proof-of-stake. Stakeholders verify transactions, whereas verification entails some cost for the stakeholders. Yet, they may also receive some rewards when they participate in transaction verification. Besides monetary considerations, the stakeholders' objectives depend on their type: Either they are well-behaving and seek verification and continuation of the blockchain or they are misbehaving and want to harm, by not validating transactions, validating them with delay, or engineering fake transactions with double-spending. In effect, these misbehaving individuals are similar to a group of voters who wants to obtain a ``negative'' voting outcome, that is, an undesirable collective decision which is detrimental to the majority. 
 Misbehaving voters are {\it byzantine}, which means that they are indistinguishable from well-behaving agents and may mimic correct behavior otherwise, except when it comes to voting.

Another concern in a blockchain governance context is very low turnout rates for voting on changes of the software and infrastructure. Voters are stakeholders and their stakes are the votes. 
Such low turnout can be explained by the costs of voting. They can consist of different types of costs: acquiring information on the issue at hand, registering for voting, being online at the time of voting, for instance. While in democracies, voting is often motivated by social norms, such social pressure is not present in the decentralized environments. 
Therefore, rational voters analyzing their probability of pivotality compared to costs of voting may abstain. Low turnout levels, however, may produce outcomes that are not in the interest of the majority of stakeholders and the voting outcome may lack credibility. On the Ethereum blockchain, turnout levels in votings on upgrade issues are 4.5\% on average, and 10\% is never reached, no matter the voting mechanism in place~\footnote{See https://www.investopedia.com/terms/o/onchain-governance.asp.}. In self-governing chains, the turnout rates in voting may drop as low as 0.12\%\footnote{See https://medium.com/wave-financial/blockchain-voter-apathy-69a1570e2af3.}. To mitigate this problem, vote delegation is proposed, in the hope that turnout levels will increase and usurping blockchain governance by few entities will become harder.

Therefore, one key question concerning blockchain management is whether stakeholders should be allowed to delegate their stake to other stakeholders, which is identical to vote delegation. Such vote delegation is part of the basic procedures of some blockchain governance, see e.g.~\cite{tezos} and the proposal in~\cite{concordium}\footnote{See page 7 in~\cite{concordium}.}. With vote delegation, well-behaving stakeholders could avoid the verification cost and earn a small return if their vote goes to other, well-behaving stakeholders. However, they incur a risk that their vote goes to misbehaving stakeholders who aim at making false transactions and disrupting the blockchain. If misbehaving agents can accumulate a majority of votes, a negative outcome occurs.

In the context of democracies---representative or direct---, suggestions are discussed that citizens should have the option to delegate their voting rights to other citizens. Citizens who often abstain could then exercise their voting rights indirectly by delegating their votes, so that  the electorate as a whole might be better represented. Yet, one drawback of vote delegation is that it may open opportunities for minorities to win with vote delegation, while the majority would prevail under conventional voting. By ``conventional voting" we mean majority voting in a direct democracy with voluntary participation but without the possibility to delegate votes.

\emph{Set-up and Results}

To delegate a vote means to transfer one's voting right to another voter. Say, for example, a voter $v_i$ delegates his/her vote to a voter $v_j$. Then, voter $v_j$ has his/her vote plus the delegated votes s/he  received---in total $2$ votes. In general, a voter can obtain and use any number of votes, if as many votes are delegated to him/her. When a voter votes, all of his/her votes count for one alternative, that is, a voter can not split his/her votes between different alternatives.

We study vote delegation with well-behaving and misbehaving agents under three assumptions. First, voting is costly for well-behaving agents. That means, if a well-behaving individual abstains or delegates his/her vote, s/he is better off than with voting as long as his/her action does not affect the voting outcome.
Second, we assume that the minority composed of misbehaving voters always votes. The rationale is that this minority is composed of determined agents who have either a strong desire to disrupt the functioning of the system, or derive a utility from enforcing their minority view that is larger than any cost of voting. The latter assumption also allows to study vote delegation under the most unfavorable circumstances. Third, we consider a ``one voter one vote" model.\footnote{Most of the blockchains allow that the voting weight of an agent is equal to his stake. We can apply our results to such weighted voting schemes if the delegation probability of the vote is proportional to the size of the stake. Approximately, this holds in our model. Details are available upon request.}

We compare vote delegation with conventional voting. In vote delegation, the well-behaving agents' decision to delegate depends on an assessment of whether to vote or to abstain and whether their vote delegation would allow misbehaving agents to obtain a majority of votes. Typically, some fraction of well-behaving agents will delegate, while the rest will vote. Since preferences are private information, delegated votes go uniformly at random to either those well-behaving agents who vote or the misbehaving agents who (always) vote. 
We model the total number of well-behaving agents as a Poisson random variable. This assumption simplifies the analysis considerably and also models the nature of uncertain large elections better. With conventional voting, well-behaving agents only compare voting with abstention and the risk that misbehaving agents obtain a majority.

 We provide three insights: First, if the number of misbehaving voters is high, both voting methods fail to deliver a positive outcome. Second, if the number of misbehaving voters is moderate, conventional voting delivers a positive outcome, while vote delegation fails with probability one. Third, with numerical simulations, we show that if the number of misbehaving voters is low, delegation delivers a positive outcome with a higher probability than  conventional voting. 
 
 Formally, we find that for any cost of voting $c$, there are thresholds $f^*(c)$ and $n^*(f)$ such that for any number of misbehaving voters $f$ above $f^*$ and an expected number of well-behaving agents above $n^*$, misbehaving voters will have the majority of votes and will win. This means that if the cost of voting is close to zero, for example, there must be many misbehaving voters and the total number of voters must be of order $f^2$, so that misbehaving voters can win. 
 
 The intuition for the non-monotonicity in the result is the following. If the number of misbehaving voters is low, delegation greatly benefits well-behaving agents who vote and thus delegation improves the chances of well-behaving agents to win. This, in turn, provides incentives to delegate. Under conventional voting, more well-behaving agents would abstain to save costs.
If the number of misbehaving voters is larger, delegation becomes quite risky, since these voters may raise the chances of misbehaving voters to win. Hence, the beneficial effects of delegation declines and conventional voting, which forces agents either to vote or to abstain, becomes dominant, since well-behaving agents have greater incentives to vote.
 
 The difference between abstention and delegation is that abstention has no impact on the voting outcome, while delegation affects the probability which alternative will be chosen---even if delegation is random---as soon as the expected number of well-behaving and misbehaving agents participating in voting differs, as the vote always goes to \textit{someone}. Individuals thus can marginally improve (or decrease) the chances that the preferred alternative wins by delegation, depending on the mix of well-behaving and misbehaving agents who vote. Hence, the individual calculuses between voting and abstention and between voting and delegation differ. For instance, in all equilibria with delegation, abstention is strictly dominated by delegation.
 
 Throughout the paper, we assume that the number of misbehaving voters, $f$, is fixed and known to well-behaving agents. This assumption can be relaxed, especially in the case of large elections, because the main result is very robust for any number of misbehaving voters from some point on. For moderate values of $f$, however, the beliefs of well-behaving agents about this value becomes more important. Our result from Proposition~\ref{moderate_values} with lower values of $f$ would still hold if this probability distribution had high concentration around the expected value. If the voters overestimate $f$ for instance, and they think it is above the threshold, nobody will vote and misbehaving voters will win with probability one. If the voters underestimate $f$, and think it is below the threshold, some of them will vote in the equilibrium and they will win with some positive probability. Therefore, our insights on equilibria solutions are still useful, depending on voters' beliefs.

 \emph{Applications}
 
 Our results have immediate implications for blockchains, i.e. that vote delegation should only be allowed if it is guaranteed that the absolute number of misbehaving agents is below a certain threshold. Otherwise, the risk for negative outcomes increases. Our results can also help assess the performance of vote delegation in democracy, a form that is known as ``liquid democracy". Indeed, for a liquid democracy, our result is worst-case result when delegating agents cannot trust those to whom they delegate. We can view misbehaving voters as a determined minority who will vote, no matter the costs. In the same setting, the well-behaving agents can be viewed as majority voters, who incur a cost and/or analyze rationally. If there is a sufficiently large minority in absolute terms who is determined to engage in its cause, vote delegation can lower the likelihood that the majority wins.

The paper is organized as follows: In Section~\ref{sec: rel_lit}, we discuss the related literature. In Section~\ref{sec:model}, we introduce our model. In Section~\ref{sec:analysis}, we analyze the equilibria and state our main result. In Section~\ref{sec: comp}, we compare the performances of vote delegation and conventional voting. Section~\ref{sec:conclusion} concludes. The proofs are in Appendix~\ref{app:proofs}.

\section{Related Literature}\label{sec: rel_lit}
Fundamental work on proxy voting and delegative democracy has been done by \cite{Tullock1967,Tullock1992} and \cite{Miller}. \cite{AlgerDanProxyVoting}, e.g., further developed this seminal work on proxy voting. He compares the performance of proxy voting to the performance of two existing voting systems in a costless direct democracy with well-informed voters. He further finds that the simple proxy voting suggested by \cite{Tullock1967} results in the best possible representation in costless voting. 

The rational voting model is discussed in~\cite{palfrey1983strategic} and~\cite{ledyard1984pure}. Vote delegation in a network and its dangers are studied in~\cite{liquid_fluid} and~\cite{liquid_algorithmic}. From an algorithmic perspective,~\cite{liquid_algorithmic} find that there is a delegation procedure that outperforms direct voting, depending on the information that the voters have. In our paper, we focus on vote delegation from a game-theoretic perspective, when preferences of agents are private information.~\cite{rational_delegation} study vote delegation in a directed graph, but do not study the case where voters aim to maximize the chance that their type wins. In our case, agents want their type to win. That is, all misbehaving voters vote for the same undesirable alternative and all well-behaving agents vote for the same desirable alternative.~\cite{blockchain_voting} study weighted voting on blockchain and consider delegation of stakes as well. 
Finally, the  literature on Poisson games started with~\cite{poisson}. In the costly voting setting,~\cite{taylor-yildirim-1} justify the use of Poisson games, which makes the analysis of pivotal probabilities easier. This setting is particularly useful when the size of the electorate is large.

Vote delegation can  take place over several voting rounds, as discussed in~\cite{casella}. In such settings, any pair of voters $(v_i,v_j)$ can write a contract about a vote exchange. A voter $v_i$ gives a vote to a voter $v_j$ in round $r$ in exchange for a vote in round $w$. So much freedom regarding vote delegation can entail an attack, even with a single misbehaving voter. A misbehaving voter can choose a voting round $t$ in the future. If $t$ is larger than half the number of all voters, the misbehaving voter can engineer vote exchange contracts with the majority of the other voters, such that at voting round $t$, s/he has the majority of votes, and thus compromises the entire system. This simple yet efficient attack called ``$t$-period attack" shows that a free intertemporal vote exchange can be very dangerous. In this paper, we explore the performance of vote delegation when there is a single delegation round.  

Another related strand of voting literature is about so-called \textit{sybil} (fake or duplicate) \textit{voters}. For example, see \cite{Meir2020SybilResilientSC} and \cite{sybilshahaf}. \cite{Meir2020SybilResilientSC} study voting procedures when honest voters are split into active and passive voters and all sybil voters are assumed to be active voters. \cite{Meir2020SybilResilientSC} study vote delegation when voters are supposed to pick a position on a line. Only a small number of voters is active and any passive voter can delegate his/her vote. The authors assume that delegated votes go to the nearest active voter on the line. This setting is quite different from our setting. The only similarity is the presence of honest and sybil voters.

Representative democracy models have been studied recently by \cite{PIVATO202052} and \cite{ijcai2019-1}. \cite{PIVATO202052} study a model where voters can choose any legislator as representative. Hence, a legislator can represent any number of voters, which will be his/her voting weight. In the actual voting all legislators vote and votes are counted according to the weights. In their model, \cite{PIVATO202052} show that for large elections, the voting outcome of the legislator's votes is the same as if all voters voted directly. \cite{SohVoutsa} uses this model and studies other forms of voting, such as \textit{Weighted Approval Voting} and \textit{Majority Judgement}. He shows that similar to \textit{Weighted Voting}, the voting outcome is the same for large elections as with direct voting.

Unlike the literature on liquid democracy, \cite{ijcai2019-1} introduce \textit{Flexible Representative Democracy}, a new model which studies a mixture of representative and direct democracy where first, a set of experts/representatives is elected by the voters for an entire term. When voting on an issue, a voter faces the decision how to allocate his/her vote among a subset of the representatives. In addition, voters can also vote directly. Comparing this model to our model, we note that random uniform delegation can be achieved by distributing the voting power of a voter uniformly among all representatives.

Furthermore, in costly voting models, voters analyze whether to vote and occur cost based on their pivotality. This measure of ``voting power"  is calculated by the Banzhaf index. \cite{zhang2020power} develop a novel index which they call \textit{Delegative Banzhaf Index}, which also measures the power of delegators.

\section{Model}\label{sec:model}
We consider a society consisting of well-behaving and misbehaving agents. In our setting, there is a good and a bad alternative. A well-behaving agent, if s/he votes at all, will vote for the good alternative and hence incur a cost $0< c\leqslant 1$. A misbehaving agent will always turn out and vote for the bad alternative to harm the system. We assume that misbehaving voters do not have any cost of voting. This can be justified by assuming that misbehaving voters are a determined minority who will vote, no matter the cost. This is why we normalize their cost to $0$. The assumption of Poisson games is that the total number of well-behaving agents $N$ is distributed as a Poisson random variable with parameter $n$,  where $n$ is some positive real number. Moreover, we assume that the number of misbehaving voters $f\in \mathbb{N}$ is common knowledge.

Each voter has the same strategy set consisting of voting and delegation. Delegation means that the vote of the delegating voter goes to some other voter. We consider a totally mixed Nash equilibrium solution concept, where well-behaving agents randomize between voting and delegating. 

Moreover, we consider a symmetric Bayesian Nash equilibrium solution concept, that is, all voters have the same probability of delegating. Let $\gamma \in [0,1]$ denote the probability of delegation, i.e., a well-behaving agent delegates his/her vote with probability $\gamma$ and votes with probability $1-\gamma$. 
A value of $\gamma$ characterizes an equilibrium if well-behaving agents are indifferent between delegating and voting.

The procedure of delegation is performed by the pooling of delegated votes and giving them uniformly at random to those who are voting. We abstract from the network structure of the vote delegation process, since the identities of the misbehaving voters are not known. Further, we assume that delegation has no cost.

From the decomposition property\footnote{See~\cite{poisson}.} of Poisson games, we have that $D$, the number of well-behaving agents who are \textit{delegating}, is distributed as a Poisson random variable with parameter $n\gamma$. On the other hand, $V$, the number of well-behaving agents who are \textit{voting}, is distributed as a Poisson random variable with parameter $n(1-\gamma)$. This means that $D$ votes are delegated to a group of $V+f$ voters consisting of the remaining $V$ well-behaving voters and $f$ misbehaving voters. Note that the random variables $D$ and $V$ are mutually independent.
Let $h$ denote the number of votes that are delegated to the well-behaving voters. Then remaining $D-h$ votes are delegated to the misbehaving voters. The following figure illustrates the setting.
\begin{figure}[htbp]
\centering
\begin{tikzpicture}
\draw (1,-2.5) circle [radius=0.7] node {$D$};
\draw (0,0) circle [radius=0.7] node {$V$};
\draw (2,0) circle [radius=0.7] node {$f$};
\draw[thick,->] (0.9,-1.8) -- (0,-0.7) node[midway, left] {$h$};
\draw[thick,->] (1.1,-1.8) -- (2,-0.7) node[midway, right] {$D-h$};
\end{tikzpicture}
\caption{$h$ out of $D$ votes are delegated to $V$ well-behaving voters and $D-h$ votes are delegated to $f$ misbehaving voters.}
\end{figure}
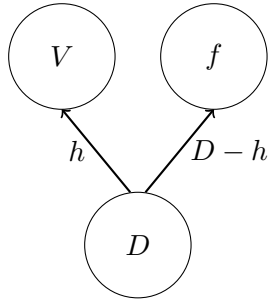

Note that by the environmental equivalence property\footnote{See~\cite{poisson}.} of Poisson games, we can take the outsider view of an additional agent, without having to readjust the parameters of the Poisson distribution. Let us distinguish between two cases for this additional agent: Either the agent is voting and hence is increasing the number of voting well-behaving agents from $V$ to $V+1$ or the voter is delegating and hence increasing the number of the delegating agents' group from $D$ to $D+1$. In the first case, the random variable $h$ follows the binomial distribution with parameters $D$ and $\frac{V+1}{V+1+f}$, because $h$ out of $D$ votes are delegated to a group of $V+1$ well-behaving voters out of $V+1+f$ voters. In the latter case, i.e., if the additional agent is delegating,  $h$ follows the binomial distribution with parameters $D+1$ and $\frac{V}{V+f}$, because now $h$ out of $D+1$ votes are delegated to $V$ well-behaving voters out of $V+f$. 

For later analysis, let us define the function
\begin{align*}
  g(x,y):=\begin{cases}
               1 & \text{if } x>y, \\
               \frac{1}{2} & \text{if } x=y,\\
               0 & \text{if } x<y,
            \end{cases}    
\end{align*}
where $g(x,y)$ denotes the gain for well-behaving voters if they receive $x$ votes, and misbehaving voters receive $y$ votes. If well-behaving voters have more votes, the gain is $1$, whereas the gain is $0$ if misbehaving voters have the majority of votes. Ties are broken by a fair coin toss, hence the expected gain of $\frac{1}{2}$ when $x$ and $y$ are equal.

\section{Vote Delegation}\label{sec:analysis}

\subsection{Main Result}

We start our analysis by the equilibrium indifference condition. The equilibrium indifference condition consists of equating the cost of voting $c$ with the difference of the expected utility of voting and the expected utility of delegating, that is, the expected utility gain from voting. We can write,
\begin{equation*}
    c = \mathbb{E}[U(\text{voting})]- \mathbb{E}[U(\text{delegating})].
\end{equation*}
The expected utilities are calculated in the following way: First, we need to sample $D$ and $V$, using the probability mass function of the Poisson distribution with parameter $n\gamma$, resp. $n(1-\gamma)$. Then, based on the sampling, we sample the number of votes $h$ that well-behaving voters are being delegated using the binomial distribution with the parameters specified above. In the case where the additional agent is voting, we have a size of $V+1$ of the well-behaving voter group. Hence, the well-behaving voters have in total $V+1+h$ votes versus $f+D-h$ misbehaving votes, which will be used to calculate the gain function $g$. Similarly, in the case where the additional agent is delegating, we have a size of $D+1$ of the delegating group. The well-behaving voters will have a total of $V+h$ votes versus $f+D+1-h$ misbehaving votes. The following equation explicitly states the indifference condition for the additional agent: 

\begin{equation}\label{eq: equicond}
\begin{split}
c&= \sum_{D=0}^\infty \sum_{V=0}^\infty \frac{(n\gamma)^D}{e^{n\gamma}D!} \frac{(n(1-\gamma))^V}{e^{n(1-\gamma)}V!} \times \\ &\quad \times \bigg[ \sum_{h=0}^D \binom{D}{h} \left(\frac{V+1}{V+1+f}\right)^h \left(\frac{f}{V+1+f}\right)^{D-h} g(V+1+h, f+D-h) \\
&\quad - \sum_{h=0}^{D+1} \binom{D+1}{h} \left(\frac{V}{V+f}\right)^h \left(\frac{f}{V+f}\right)^{D+1-h} g(V+h, f+D+1-h) \bigg].
\end{split}
\end{equation}
Let us elaborate on the right-hand side of this equation. First, we call the right-hand side of~\eqref{eq: equicond} from now on $\xi _{n,f}(\gamma)$. The first two sums, together with the Poisson probability mass functions, account for the sampling of $D$ and $V$. Whereas the other two sums, together with the binomial probability mass functions, account for the sampling of $h$.

Next, we state our main result.
\\
\begin{theorem}\label{thm:f_large_no_sol}
For any $c>0$, there exists $f^*(c)\in \Theta(\frac{\delta^2}{c^2})$ and $n^*(f)\in \Theta(\frac{f^2}{\delta^2})$, for any $\delta\in \mathbb{N}$, so that, for all $f\geq f^*(c)$ and $n\geq n^*(f)$, well-behaving agents lose.
\end{theorem}

\begin{proof}
See Appendix~\ref{app:proofs}.
\end{proof}

Well-behaving agents losing is equivalent to the fact that there exists no solution to the equation~\eqref{eq: equicond}. In particular, the right-hand side of equation~\eqref{eq: equicond} is strictly smaller than $c$. Or, equivalently, every well-behaving agent is delegating, because the cost of voting is larger than the expected benefits for voting, and hence misbehaving voters  have the majority (in fact, all) of the votes. From the theorem, we see that lowest value for $f$ is for constant $\delta$, and it is from the set of functions $\Theta\left(\frac{1}{c^2}\right)$. In this case, however, $n$ needs to be large as a function of $f$, namely ${f^2}$. Verbally, the theorem states that vote delegation is dangerous if there are sufficiently many misbehaving voters and sufficiently many well-behaving agents.
The intuition behind this result is the following. If the number of well-behaving voters is even slightly smaller than the number of misbehaving voters, then with high probability, the number of delegating agents is large enough to guarantee that well-behaving agents will lose. That is, costly voting is not a dominant strategy. If the number of well-behaving voters is much bigger than the number of misbehaving voters, then for any number of delegating agents, well-behaving voters will win with high probability. On the other hand, if the number of well-behaving voters is moderately larger than the number of misbehaving agents, then the number of delegating agents is large enough to guarantee that well-behaving agents win with high probability. 
That is, in neither of the latter two cases a well-behaving player is motivated to vote. Last, the probability that the number of well-behaving voters is approximately equal to the number of misbehaving voters is sufficiently small. It follows from a large enough expected number of misbehaving voters and from the Poisson random variable properties.

\subsection{Winning Probability and Social Welfare}
To assess the efficiency of a voting rule, we look at two values. First, we consider the probability that well-behaving agents win in equilibrium under a specific voting rule. Second, we consider the value of per-capita social welfare, which consists of the expected benefits of the whole (well-behaving) society minus the costs incurred by voting.  
The probability $p$ that the well-behaving agents win is calculated by the following formula:
\begin{align*}
    p(n,f,\gamma) &= \sum_{D=0}^\infty \sum_{V=0}^\infty \sum_{h=0}^D \frac{(n\gamma)^D}{e^{n\gamma}D!} \frac{(n(1-\gamma))^V}{e^{n(1-\gamma)}V!} \binom{D}{h} \left(\frac{V}{V+f}\right)^h \left(\frac{f}{V+f}\right)^{D-h} \\
    &\qquad g(V+h, f+D-h).
\end{align*}

We note that $D$ and $V$ are sampled and then, given $D$, $h$ is sampled. Therefore, we have $V+h$ votes for the right alternative and $f+D-h$ votes for misbehaving voters. This implies that well-behaving agents win with certainty if $V+h>f+D-h$ and they win with probability $\frac{1}{2}$ when $V+h=f+D-h$. The probability that well-behaving agents win is described by the $g$ function introduced in the last section. 

We calculate per-capita social welfare as the difference between the probability of winning and the voting costs. The former represents per capita expected utility. Per-capita welfare is thus given by
\begin{align*}
    W(n,f,\gamma,c) &= \sum_{D=0}^\infty \sum_{V=0}^\infty r(D+V)\sum_{h=0}^D \frac{(n\gamma)^D}{e^{n\gamma}D!} \frac{(n(1-\gamma))^V}{e^{n(1-\gamma)}V!} \binom{D}{h} \left(\frac{V}{V+f}\right)^h \left(\frac{f}{V+f}\right)^{D-h}\\ 
    &\qquad \bigg((D+V)g(V+h, f+D-h)-Vc\bigg),
\end{align*}

where the function $r$ is defined as follows: 

\begin{align*}
r(x):=\begin{cases}
               \frac{1}{x} & \text{if } x>0, \\
               0 & \text{if } x=0.
        \end{cases}
\end{align*}

Note that the term $-Vc$ in the above formula is the cost spent on voting by well-behaving voters, and it is subtracted from the total benefits for the society. The latter is captured by the term $(D+V)g(V+h, f+D-h)$. 

\section{Comparison with Conventional Voting}\label{sec: comp}

\subsection{Conventional Voting}

With conventional voting, there is no option to delegate one's voting right. Thus, the voters' strategy sets only consist of voting and abstaining. We look for symmetric Bayesian Nash equilibria under conventional voting. Let $\alpha \in [0,1]$ be the probability of voting. To determine $\alpha$,  we next derive the indifference condition between voting and abstention. We have to consider the cases where an additional vote would impact the probability that well-behaving agents win. Then, we have to equate the cost of voting, $c$, with the difference between the expected utilities of voting and abstaining, that is
\begin{equation*}
    c = \mathbb{E}[U(\text{voting})]- \mathbb{E}[U(\text{abstaining})].
\end{equation*}
This indifference can only hold if well-behaving agents have either $f$ or $f-1$ votes. In the case of $f$ votes, there is a draw ($f$ versus $f$). Then, one additional vote from well-behaving agents will win and creates a utility gain of $\frac{1}{2}$. In the other case, where well-behaving agents have exactly $f-1$ votes, one additional vote by well-behaving agents would turn this loss into a draw, which again yields a utility gain of $\frac{1}{2}$. These expected utilities are equated with the cost of voting and we obtain the following indifference relation between voting and abstaining:
\begin{align}\label{eq: equicond_base}
    c = \frac{1}{2}\frac{(n\alpha)^f}{e^{n\alpha}f!} + \frac{1}{2}\frac{(n\alpha)^{f-1}}{e^{n\alpha}(f-1)!}.
\end{align}
For a given probability of voting $\alpha$, the probability of the well-behaving agents winning in the baseline game is calculated by the following formula:
\begin{align*}
    q(n,f,\alpha) = \sum_{k=0}^\infty \frac{(n\alpha)^k}{e^{n\alpha}k!}g(k,f).
\end{align*}
The per-capita social welfare is calculated as  the difference between the probability of winning, measuring the expected benefits, and the cost of voting. To calculate per-capita social welfare, we first sample the total number of citizens, denoted by $N$ and then, out of those $N$ individuals, $k$ individuals will vote with probability $\alpha$. That is, $D$ is sampled as a Binomial random variable with parameters $N$ and $\alpha$, which leads to the following formula:
\begin{align*}
    W(n,f,\alpha,c) = \sum_{N=0}^{\infty}\frac{n^N}{e^nN!}r(N)\sum_{k=0}^{N}{N \choose k}\alpha^k(1-\alpha)^{N-k}(Ng(k,f)-kc),
\end{align*}

where the $r(N)$ term has been introduced above and measures the per-capita benefit when there are $N$ well-behaving agents in total. The term $Ng(k,f)$ stands for the total benefits, while $-kc$ stands for the costs incurred by $k$ voters voting. 

\subsection{Comparisons}

In this section, we compare the performance of vote delegation with conventional voting and we do this in two ways: the probability that the right alternative is winning and per-capital social welfare. 

From the proof of Theorem~\ref{thm:f_large_no_sol} and by taking $\delta=1$, we obtain the lower bound threshold on $f^*(c)$ for vote delegation, above which no well-behaving agent votes. We can compare this to the corresponding threshold for conventional voting that has been derived in~\cite{AV}. This leads to the following proposition. 

\begin{proposition}\label{moderate_values}
There are constants $0 < t_1 < t_2$, such that 
\begin{itemize}
    \item [(i)] If $f > t_2 \frac{1}{c^2}$, misbehaving voters will win with probability 1 under vote delegation and under conventional voting.
    \item [(ii)] If $f\in[t_1 \frac{1}{c^2}, t_2 \frac{1}{c^2}]$, the probability that well-behaving voters win with vote delegation is zero, while the probability that well-behaving voters win with conventional voting is positive.
\end{itemize}
\end{proposition}

Hence, with the above proposition, we obtain two insights: First, if the number of misbehaving voters is high, both voting methods fail to deliver a positive outcome. Second, if the number of misbehaving voters is moderate, conventional voting delivers a positive outcome, while vote delegation fails with probability one. Hence, in such cases, conventional voting performs better than vote delegation,  

Next, we numerically compare the performance of two voting rules for small values of $f$ and $n$. The analysis around Theorem~\ref{thm:f_large_no_sol} requires a large electorate, that is, $n$ and $f$ need to be large enough numbers for the result to hold. Theorem~\ref{thm:f_large_no_sol} tells us that there are no mixed equilibria solutions if $n$ and $f$ are large enough. 

By numerical simulations, we find that there are mixed equilibria solutions if $n$ and $f$ are small enough. We illustrate this by the following example. The cost of voting $c$ is equal to $0.14$. The expected number of well-behaving agents, $n$, is equal to $30$. Approximate numerical solutions show that we have two different equilibria solutions for any $f>1$, but $f$ not too large. In Table 1,  the probabilities $p_1$ and $p_2$ represent the equilibrium probabilities of well-behaving agents winning in the delegation game. In Table 2, the values $q_1$ and $q_2$ represent the equilibrium probabilities of well-behaving agents winning in the conventional voting game. In both tables, $W_1$ and $W_2$ are the per-capita social welfare values for the corresponding voting game.

\begin{figure}[htbp]
\begin{floatrow}
\ffigbox{
  \includegraphics[width = 0.5\textwidth]{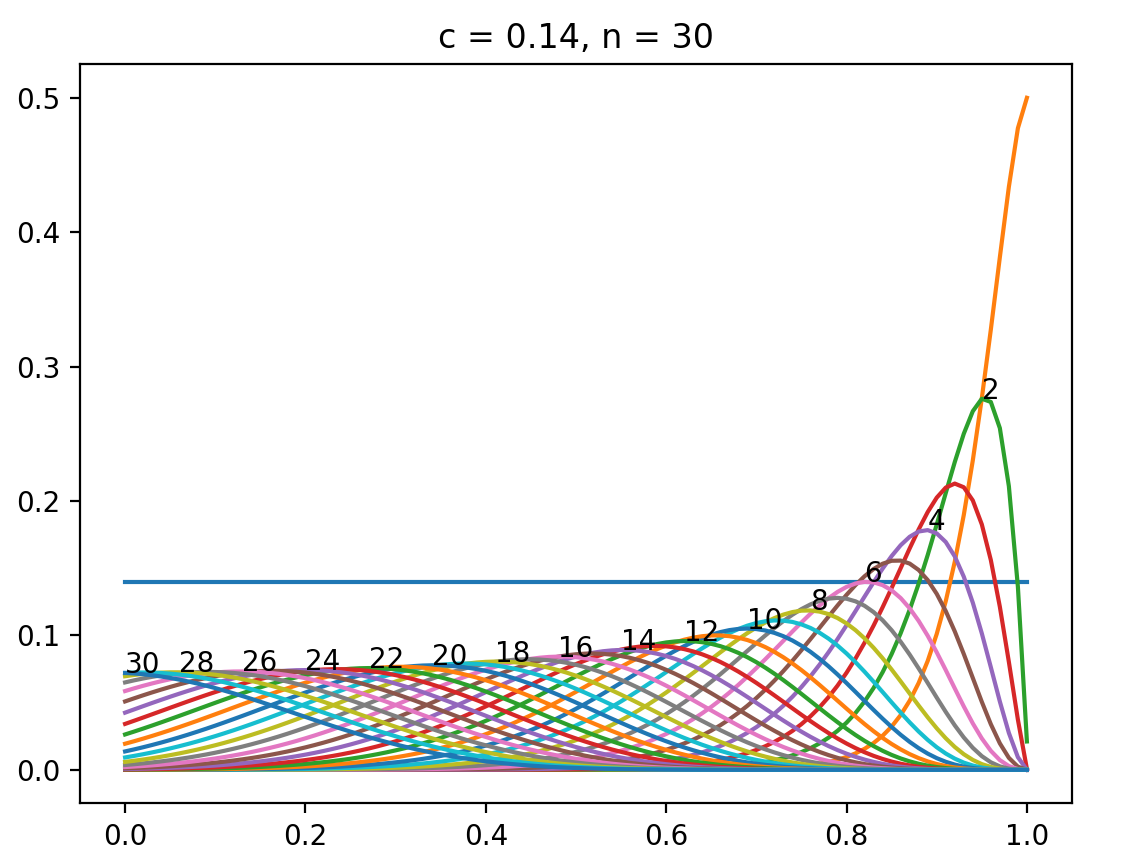}
}{%
  \caption{Delegation game: The $x$-axis  represents values of $\gamma$ and the $y$-axis displays expected utility gains from voting, with the blue horizontal line representing cost. The graphs represent the right hand side (RHS) of equation (\ref{eq: equicond}) for different values of $f$.}%
}
\capbtabbox{
     \begin{tabular}{c|c|c||c|c} 
    \hline
    \hline
    $f$ & $p_1$& $p_2$ & $W_1$& $W_2$ \\
    \hline
    \hline
       1 &   0.84   &   -      &   0.82   &   -     \\
       2 &   0.76   &    0.08      &   0.74   &   0.07   \\
       3 &     0.71      &  0.10       &   0.69   &   0.09  \\
       4 &   0.63     &     0.14  &   0.61   &   0.13  \\
      5 &    0.57    &   0.21      &    0.54  &   0.19  \\
    6-20 &    -     &    -     &   -  &   - \\
    \end{tabular}
}{%
  \caption{Equilibrium probabilities $p_1,p_2$ of well-behaving agents winning with delegation and per-capita social welfare values of the corresponding equilibria states $W_1, W_2$ for $c = 0.14$ and $n=30$.}%
}
\end{floatrow}
\end{figure}

\begin{figure}[htbp]
\begin{floatrow}
\ffigbox{
  \includegraphics[width = 0.5\textwidth]{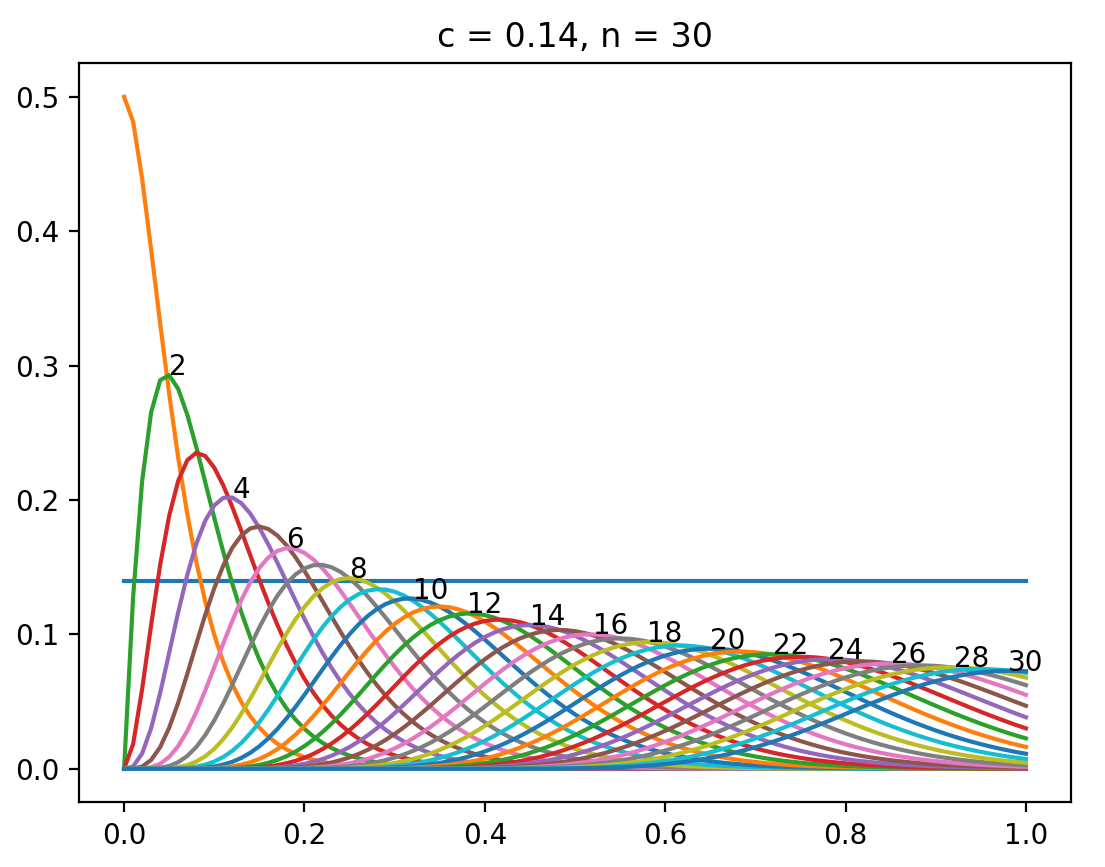}
}{%
  \caption{Conventional game: The $x$-axis  represents values of $\alpha$ and the $y$-axis displays expected utility gains from voting, with the blue horizontal line representing cost. The graphs represent the RHS of equation (\ref{eq: equicond_base}) for different values of $f$.}%
}
\capbtabbox{
     \begin{tabular}{c|c|c||c|c} 
    \hline
    \hline
    $f$ & $q_1$& $q_2$ & $W_1$& $W_2$ \\
    \hline
    \hline
       1 &   0.80   &   -      &   0.79   &   -     \\
       2 &   0.74   &    0.02      &   0.73   &   0.02   \\
       3 &     0.74      &  0.04       &   0.72   &   0.03  \\
       4 &   0.66     &     0.07  &   0.64   &   0.06  \\
      5 &    0.63    &   0.13      &    0.61  &   0.12  \\
     6 &    0.61    &   0.19      &    0.58  &   0.17  \\
      7 &    0.55    &   0.24      &    0.51  &   0.21  \\
      8 &    0.45    &   0.32      &    0.41  &   0.29  \\
      9-30 &    -     &    -     &   -  &   - \\
    \end{tabular}
}{%
  \caption{Equilibrium probabilities $q_1,q_2$ of well-behaving agents winning in the conventional voting game and per-capita social welfare values of the corresponding equilibria states $W_1, W_2$ for $c = 0.14$ and $n=30$.}%
}
\end{floatrow}
\end{figure}

Extensive numerical calculations for different parameter values correspond to the pattern shown in the tables. They suggest that the two equilibrium probabilities of well-behaving agents winning in the delegation game are higher than the probabilities in the conventional voting game for sufficiently low values of $f$. 
For moderate values of $f$, however, the conventional voting game yields higher probabilities that the correct alternative is chosen and vote delegation may even prevent the correct alternative from having any chance to win. Finally, for high values of $f$, misbehaving voters win with certainty in both, the delegation and the conventional voting games. The latter observation is in line with our main result. 

We address the special case $f=1$. In this case, with vote delegation, we have a totally mixed equilibrium for a large enough value of $c$, and the equilibrium is unique. \\

\begin{proposition}\label{lem:f1gamm1}
For $f=1$, equation~\eqref{eq: equicond} has a solution for any $c\in [\frac{1}{e^n} (\frac{1}{2}+\frac{n}{2}+\frac{n^2}{12}),\frac{1}{2}]$. 
\end{proposition}

\begin{proof}
See Appendix~\ref{app:proofs}.
\end{proof}

We note that for $n$ arbitrarily large, the left endpoint of the interval in the proposition converges to $0$. Therefore, the interval of $c$ for which there is a solution converges to $[0,\frac{1}{2}]$ and $\frac{1}{2}$ is a natural upper bound for the cost of voting.

For the conventional voting game,  we also obtain a totally mixed equilibrium for a large enough value of $c$, and the equilibrium is unique. In particular, we obtain the following result:

\begin{corollary}\label{prop: base}
For any $f\geq 2$, equation~\eqref{eq: equicond_base} has a solution for any $c \in [0, \frac{1}{e^{\sqrt{f(f-1)}}} \left(\frac{\sqrt{f(f-1)}^f}{f!} + \frac{\sqrt{f(f-1)}^{f-1}}{(f-1)!} \right)]$.
For $f=1$, equation~\eqref{eq: equicond_base} has a solution for any $c \in [\frac{n+1}{2 e^n},\frac{1}{2}]$.
\end{corollary}

\begin{proof}
See Appendix~\ref{app:proofs}.
\end{proof}

In the case $f\geq 2$, we note that the right endpoint of the interval converges towards $0$ as $f$ grows and reaches the critical value above which no well-behaving individual will vote anymore.
In the case $f=1$, the left endpoint of the interval converges to $0$ if $n$ grows unboundedly.  


\section{Conclusion} \label{sec:conclusion}
We have studied the role of vote delegation in the costly voting setting. In particular, we showed that with malicious parties, vote delegation is a risky procedure if the number of misbehaving voters is not low. 
However, for a low number of misbehaving voters, we showed that vote delegation dominates conventional voting, i.e., it  implements the right alternative with a higher probability than conventional voting. Overall, our results suggest that one should be cautious with the implementation of vote delegation. 

Our setting of costly voting is orthogonal to the information acquisition setting studied in the literature. We believe the cost-saving aspect of vote delegation makes it worth studying further. Our analysis can be extended in various ways. First, we could study caps on the number of delegated votes, that is, every voter is only allowed to cast at most a constant maximum number of votes. Second, we could assume that there is a number of committed well-behaving agents who will vote no matter the costs. Note that this case is not equivalent to considering fewer misbehaving voters, because having some number of well-behaving agents voting with certainty changes the probability that a vote will be delegated to a misbehaving voter.

\subsection*{Acknowledgements}
This research was partially supported by the Zurich Information Security and Privacy Center (ZISC).

\bibliographystyle{apalike}
\bibliography{references}

\clearpage

\appendix
\section{Proofs} \label{app:proofs}

\begin{proof}[Proof of Theorem \ref{thm:f_large_no_sol}]
We analyze $\xi_{n,f}(\gamma)$, the right-hand side of equation~\eqref{eq: equicond}.
To make notation easier, we define the function
\begin{align*}
    G_f(D,V):= \sum_{h=0}^D \binom{D}{h} \left(\frac{V+1}{V+1+f}\right)^h \left(\frac{f}{V+1+f}\right)^{D-h} g(V+1+h, f+D-h) \\
    \qquad - \sum_{h=0}^{D+1} \binom{D+1}{h} \left(\frac{V}{V+f}\right)^h \left(\frac{f}{V+f}\right)^{D+1-h} g(V+h, f+D+1-h).
\end{align*}
Note that by the property of a probability distribution and by the definition of $g$, we have  $|G_f(D,V)| \leqslant 1$.
Our goal is to show that $c > \xi_{n,f}(\gamma)$ for any $\gamma \in [0,1]$, for large $f$.
Equivalently, we prove that the value $\xi_{n,f}(\gamma)$ is very small. That is, we show it is close enough to zero for large enough $f$.

First recall the definition
\begin{equation*}
    \xi_{n,f}(\gamma) = \sum_{D=0}^\infty \sum_{V=0}^\infty \frac{(n\gamma)^D}{e^{n\gamma}D!} \frac{(n(1-\gamma))^V}{e^{n(1-\gamma)}V!} G_f(D,V).
\end{equation*}

Next, we derive upper bounds on $G_f(D,V)$.
Let us rewrite 
\begin{align}
    G_f(D,V) & = \sum_{h=\lfloor \frac{f+D-V+1}{2}\rfloor}^D \binom{D}{h} \left(\frac{V+1}{V+1+f}\right)^h \left(\frac{f}{V+1+f}\right)^{D-h} \label{eq: a1} \\ \label{eq: a11}
    & \qquad + \frac{1}{2}\binom{D}{\frac{f+D-V-1}{2}} \left(\frac{V+1}{V+1+f}\right)^\frac{f+D-V-1}{2} \left(\frac{f}{V+1+f}\right)^{D-\frac{f+D-V-1}{2}} \\ \label{eq: a2}
    &\qquad - \sum_{h= \lfloor \frac{f+D+3-V}{2}\rfloor}^{D+1} \binom{D+1}{h} \left(\frac{V}{V+f}\right)^h \left(\frac{f}{V+f}\right)^{D+1-h}\\ \label{eq: a22}
    &\qquad - \frac{1}{2}\binom{D+1}{\frac{f+D+1-V}{2}} \left(\frac{V}{V+f}\right)^\frac{f+D+1-V}{2} \left(\frac{f}{V+f}\right)^{D-\frac{f+D+1-V}{2}}.
\end{align}
The above equation holds because 
\begin{equation*}
    g(V+1+h,f+D-h)=1 \iff h>\frac{f+D-V-1}{2},
\end{equation*} and
\begin{equation*}
g(V+1+h,f+D-h)=\frac{1}{2} \iff h=\frac{f+D-V-1}{2}.    
\end{equation*}
Analogously, for $g(V+h, f+D+1-h)$, we have 
\begin{equation*}
    g(V+h,f+D+1-h)=1 \iff h>\frac{f+D+1-V}{2}
\end{equation*} and
\begin{equation*}
g(V+h,f+D+1-h)=\frac{1}{2} \iff h=\frac{f+D+1-V}{2}.    
\end{equation*}

We define 
\begin{equation}\label{eq: a12}
    a_1 = \frac{f+D-V-1}{2}, \text{ and } a_2 = \frac{f+D+1-V}{2}.
\end{equation}
Further, we define random variables $X_1, X_2$, so that $X_1$ is distributed as $Bin(D,\frac{V+1}{V+1+f})$, a binomial random variable with parameters $D$ and $\frac{V+1}{V+1+f}$, and $X_2$ is distributed as $Bin(D+1,\frac{V}{V+f})$. Then, equation~\eqref{eq: a1} is equal to the probability $P[X_1 > a_1]$ and equation~\eqref{eq: a2} is equal to $-P[X_2 > a_2]$. Similarly,~\eqref{eq: a1} +~\eqref{eq: a11} is equal to $P[X_1 > a_1]+\frac{1}{2}P[X_1 = a_1]$ and~\eqref{eq: a2} +~\eqref{eq: a22} is equal to $-P[X_2 > a_2]-\frac{1}{2}P[X_2 = a_2]$. Together we have,
\begin{equation}\label{eq: G_f_in_terms_of_probs}
    G_f(D,V) = P[X_1 > a_1]+\frac{1}{2}P[X_1 = a_1]-P[X_2 > a_2]-\frac{1}{2}P[X_2 = a_2].
\end{equation}
where $\frac{1}{2}P[X_1 = a_1]$ and $\frac{1}{2}P[X_2 = a_2]$ vanish if $a_1$, resp. $a_2$, are not integers.
Let $\delta(f), \Tilde{\delta}(f), \sigma(f)$ be some functions of $f$, determined later in the proof.
We consider the following three cases:

\begin{itemize}
    \item $V \leq f-\delta(f)$,  low value of $V$.
    \item $V \geq f+\Tilde{\delta}(f)$, intermediate value of $V$. 
    \item $V \in (f-\delta(f), f+\Tilde{\delta}(f))$, high value of $V$.
\end{itemize}

We resolve all cases in the following. 
\\\\
\textbf{Case 1. } Let $V \leq f-\delta(f)$. 
As $V$ is small, $D$ is large with high probability since $V$ and $D$ are distributed as Poisson random variables with parameters $n(1-\gamma)$ and $n\gamma$, respectively. That is, if one value is small, the other is large with high probability.

The larger $D$, the higher the chance that well-behaving voters obtain more votes. In the worst case for well-behaving voters, $V = f- \delta(f)$, i.e. $V$ is as large as possible in Case 1 and hence $D$ will be smaller with high probability than what it would be with high probability if $V$ were even lower than $f-\delta(f)$.

Let us consider equality, $V = f-\delta(f)$. Then $p$ the probability that a vote is delegated to $V+1$ well-behaving voters is
\begin{equation*}
    p = \frac{V+1}{V+1+f} = \frac{1}{2} - \frac{\delta(f)-1}{2(2f - \delta(f)+1)},
\end{equation*} 
and $a_1$, as given in~\eqref{eq: a12}, is 
\begin{equation*}
    a_1 = \frac{f+D-V-1}{2} = \frac{D+\delta(f)-1}{2}.
\end{equation*} 
We want to upper-bound the first two terms in~\eqref{eq: G_f_in_terms_of_probs}, i.e. $P[X_1 \geq a_1] < \frac{c}{3}$. First we state Hoeffding's inequality\footnote{See~\cite{Hoeffding}.} which gives us an exponential upper bound for some real-valued $\epsilon$
\begin{equation}\label{eq: hoeffding1}
    P[X_1 \geqslant D(p+\epsilon)] \leqslant \exp(-2 \epsilon^2 D).
\end{equation}
In order to make use of~\eqref{eq: hoeffding1}, we need to find $\epsilon$ first. By setting $D(p + \epsilon) = a_1$, we can solve this for $\epsilon$:
\begin{equation}\label{eq: eps}
    \epsilon = \frac{a_1}{D}-p = \frac{\delta(f)-1}{2}(\frac{1}{D} + \frac{1}{2f - \delta(f)+1})
\end{equation} 

Note that for $D > 2f$, we can upper bound $\epsilon$ to make the ensuing analysis easier,
\begin{equation}\label{eq: eps_upp_bound1}
    \epsilon \leq \frac{3\delta(f)}{4f}.
\end{equation}
By the inequality of~\cite{Mitzenmacher} for $2f < n \gamma$ we can lower-bound the probability that $D>2f$, 
\begin{equation*}
    P[D > 2f]\geq 1- 
    \frac{e^{-n\gamma}(en\gamma)^{2f}}{(2f)^{2f}}.
\end{equation*}
As $n$ is large, we see that this probability is high. That is, for any $\beta >0$ there exists $n^*$,  such that for all $n\geq n^*$, we have $$ \frac{e^{-n\gamma}(en\gamma)^{2f}}{(2f)^{2f}} \leq \beta.$$ The threshold $n^*$ is calculated by solving the above inequality. Later in the proof, we will obtain more thresholds for $n$. In the end we choose the maximum of all thresholds.
Hence we can use~\eqref{eq: eps_upp_bound1}.

Next, we make use of the upper bound in~\eqref{eq: hoeffding1}. We set $\exp(-2\epsilon^2 D) = \frac{c}{3}$ and solve it for $D$. But remember that now, we are using the upper bound on $\epsilon$,~\eqref{eq: eps_upp_bound1}:

\begin{equation*}
    \exp(-2\epsilon^2 D) = \frac{c}{3} \geq \exp(-2(\frac{3\delta(f)}{4f})^2 D).
\end{equation*}
We end up with a lower bound on $D$ which we call $D^*$,
\begin{equation}\label{lower_bound1}
    D \geq \frac{f^2}{\delta^2(f)} \frac{8}{9}\log(\frac{3}{c}) =: D^*.
\end{equation}

Then, for any $D \geq D^*$ we have $P[X_1 \geq a_1]<\frac{c}{3}$ and hence,
\begin{align}\label{eq: G<c/3_case1}
    \begin{split}
        G_f(D,V) &= P[X_1 > a_1]+\frac{1}{2}P[X_1 = a_1]-P[X_2 > a_2]-\frac{1}{2}P[X_2 = a_2] \\
        & \leq P[X_1 \geq a_1] < \frac{c}{3},
    \end{split}
\end{align}
where the last inequality precisely follows from Hoeffding's inequality.

In the last step, it remains to show that the probability that $D\geq D^*$ is high and hence~\eqref{eq: G<c/3_case1} holds.
For this, we can use the inequality from~\cite{Mitzenmacher} for $D^* < n\gamma$,
\begin{equation*}
    P[D\leq D^*] \leq \frac{e^{-n\gamma}(en\gamma)^{D^*}}{(D^*)^{D^*}}.
\end{equation*}
With $n$ large enough or by taking $$\delta(f)>\frac{2f}{\sqrt{\frac{n\gamma}{2\log(3/c)}}+1},$$ we can ensure that $D^* < n\gamma$ and hence by~\cite{Mitzenmacher},
\begin{equation*}
    P[D\geq D^*] = 1-P[D\leq D^*] \geq 1-\frac{e^{-n\gamma}(en\gamma)^{D^*}}{(D^*)^{D^*}}.
\end{equation*}

 We find parameters such that this event has a probability at least $q$, where $q$ is close to 1.

 Note that
\begin{equation*}
    \iota(n) := \frac{e^{-n\gamma}(en\gamma)^{D^*}}{(D^*)^{D^*}}
\end{equation*}
has only one extremum for $\gamma, D^*,n \neq 0$, which attains its maximum at
\begin{equation*}
    n^* = \frac{D^*}{\gamma}.
\end{equation*}
Further, note that $\iota(n) \to 0$ as $n \to \{0, \infty\}$. 
This means that for any $\beta > 0$, there are some $0<n_1<n^*<n_2$, so that for all $n<n_1$ and all $n>n_2$, we have $\iota(n) < \beta$.

Let $0<\beta<1$. We solve $\iota(n) = \beta$. We know from above that there are two solutions:
\begin{alignat}{2}
   && \frac{e^{-n\gamma}(en\gamma)^{D^*}}{(D^*)^{D^*}} &= \beta \\ \label{eq: l1}
    \iff && -n \frac{\gamma}{D^*} e^{-n \frac{\gamma}{D^*}} &= -\frac{\beta^{1/D^*}}{e}.
\end{alignat}
To solve this, we use the Lambert $W$ function. A short overview of the Lambert $W$ function is given in~\cite{lambert}. The Lambert $W$ function is the inverse function of $f(w)=we^w$. It is multi-valued and has a infinite number of branches $W_k(z)$ for $k\in \mathbb{Z}$ and $z \in \mathbb{C}$. $W_k(z)$ is complex everywhere except for $k = 0$ and $k=-1$. $W_0(z)$ is real-valued and monotone increasing for $\mathbb{R} \ni z \geq -1/e$, with the image $[-1,+\infty)$. $W_{-1}(z)$ is real-valued and monotone decreasing for real $z \in [-1/e,0)$, with the image $(-\infty,-1]$. We want to apply the Lambert $W$ function to equation~\eqref{eq: l1}. On the left hand side (LHS), we have $we^w$ for $w = -n \frac{\gamma}{D^*}$. On the RHS we have a real number that is in the interval $(-1/e,0)$, since $\beta \in (0,1)$. As the RHS is in the valid interval, we can apply two two branches $W_0$ and $W_{-1}$ to~\eqref{eq: l1}. Let $k \in \{0,-1\}$
\begin{alignat*}{2}
   \eqref{eq: l1}\iff && W_k(-n \frac{\gamma}{D^*} e^{-n \frac{\gamma}{D^*}}) &= W_k(-\frac{\beta^{1/D^*}}{e})\\
    \iff && -n\frac{\gamma}{D^*} &=W_k(-\frac{\beta^{1/D^*}}{e})\\
    \iff && n &= -\frac{D^*}{\gamma} W_k(-\frac{\beta^{1/D^*}}{e}).
\end{alignat*}
As $W_0(-1/e)=W_{-1}(-1/e) = -1$ and $W_0(0) = 0$,
\begin{equation*}
    W_0(-\frac{\beta^{1/D^*}}{e}) \in (-1,0)
\end{equation*}
and 
\begin{equation*}
    W_{-1}(-\frac{\beta^{1/D^*}}{e}) \in (-\infty,-1).
\end{equation*}
Hence, the two solutions to equation~\eqref{eq: l1} are
\begin{equation*}
    n_1 := -\frac{D^*}{\gamma} W_0(-\frac{\beta^{1/D^*}}{e}) < -\frac{D^*}{\gamma} W_{-1}(-\frac{\beta^{1/D^*}}{e}) =: n_2.
\end{equation*}
Clearly, we see that $n_1 < n^* < n_2$.

Therefore, we obtain two thresholds $n_1, n_2$ using the Lambert $W$ function, one for applying $W_0$ and the other for applying $W_{-1}$:
 \begin{equation*}
         n_{1,2} = -\frac{D^*}{\gamma} W_{0,-1}(-\frac{(1-q)^{1/D^*}}{e}).
 \end{equation*}
We choose the higher threshold. \\

\noindent
\textbf{Case 2. } Let $V \geq f + \Tilde{\delta}(f)$. Then we have to consider two cases: First, if $V$ is much larger than $f$, e.g. larger than $f + \sigma(f)$. In this case well-behaving agents win for any $D$. The second case is the one where $V$ is larger than $f +\Tilde{\delta}(f)$ and smaller than $f+\sigma(f)$.\\

Let $V > f+\sigma(f)$. We then have to distinguish between two cases: $D<\sigma(f)$ and $D>\sigma(f)$.
\begin{itemize}
    \item If $D<\sigma(f)$, then well-behaving agents win, because even if all $D$ votes are being delegated to misbehaving voters, these misbehaving voters still end up having less votes. 
    \item If $D>\sigma(f)$, we define a binomial random variable $Y$ that stands for the number of votes (out of $D$) that are delegated to well-behaving voters. The probability that a vote goes to a well-behaving voter is $p = \frac{V}{V+f}$. Note that $p$ is an increasing function in $V$. Although $V>f+\sigma(f)$, we consider the case where $V = f+\sigma(f)$. Hence we consider a lower probability $p$ than the actual $p$. Now, 
    \begin{equation*}
        p = \frac{V}{V+f} = \frac{f+\sigma(f)}{2f+\sigma(f)} = 1-\frac{f}{2f+\sigma(f)}.
    \end{equation*} 
    Note that $Y \sim Bin(D,p)$. The probability that well-behaving agents can at most create a tie is 
        \begin{align*}
        P[Y \leq \frac{D-(V-f)}{2}] &< P[Y \leq \frac{D}{2} - \frac{\sigma(f)}{2}] \\ &\leq \exp(-2 (\frac{1}{2} - \frac{f}{2f + \sigma(f)} + \frac{\sigma(f)}{2D})^2 D) \\&= \exp(-2 \sigma^2(f) (\frac{1}{4f +2\sigma(f)} + \frac{1}{2D})^2 D) \\ &= \exp(-2\sigma^2(f) (\frac{D}{(4f+2\sigma(f))^2} + \frac{1}{4f+2\sigma(f)} + \frac{1}{4D})),
    \end{align*}
   where we used the fact that $V-f > \sigma(f)$ for the first strict inequality. For the second inequality, we solved $D(p-\epsilon) = \frac{D}{2} - \frac{\sigma(f)}{2}$ and used Hoeffding's inequality. We see that the last function of the expression converges to $0$ as $f$ increases. This means again that for sufficiently high $f$, well-behaving agents win.
   
   Taking these together, we find that in the case $V > f+\sigma(f)$, well-behaving agents win for sufficiently high $f$.
   The fact that well-behaving agents win means that they have more votes than misbehaving voters, i.e. $g(\cdot, \cdot) = 1$. Hence, $G_f(D,V) = 0$ in this case, as we subtract to binomial sums which are both equal to $1$.
\end{itemize}

Let $f+\Tilde{\delta}(f) \leq V \leq f +\sigma(f)$.
\begin{itemize}
    \item Let $V = f + \Tilde{\delta}(f)$. Then, the probability $p$ that a vote is being delegated to the well-behaving voters is 
    \begin{equation*}
        p = \frac{V}{V+f} = \frac{1}{2} + \frac{\Tilde{\delta}(f)}{2(2f+\Tilde{\delta}(f))}
    \end{equation*}
    and $a_2$, as given in~\eqref{eq: a12}, is
    \begin{equation*}
        a_2 = \frac{f+D+1-V}{2} = \frac{D+1+\Tilde{\delta}(f)}{2}.
    \end{equation*} 
    We want to upper-bound the following: $P[X_2 \leq a_2] < c/3$. Again we recall Hoeffding's inequality for some real-valued $\epsilon$,
    \begin{equation}\label{eq: hoeffding2}
        P[X \leqslant D(p-\epsilon)] \leqslant \exp(-2 \epsilon^2 D).
    \end{equation} 
     To use~\eqref{eq: hoeffding2}, we first need to find $\epsilon$. By setting $D(p - \epsilon) = a_2$, we can solve $\epsilon$: 
    \begin{equation}\label{eq: eps2}
        \epsilon = p - \frac{a_2}{D} = \frac{\Tilde{\delta}(f)}{4f + 2 \Tilde{\delta}(f)}- \frac{\Tilde{\delta}(f)+1}{2D}. 
    \end{equation}
    
Note that for $D > 8f$, we can upper-bound $\epsilon$ as we did in Case 1,
\begin{equation}\label{eq: eps_upp_bound2}
    \epsilon \leq \frac{\Tilde{\delta}(f)}{4f}.
\end{equation}
By the inequality from~\cite{Mitzenmacher}, for $8f < n \gamma$, we can lower-bound the probability of the event $D>8f$:
\begin{equation*}
    P[D > 8f]\geq 1- 
    \frac{e^{-n\gamma}(en\gamma)^{8f}}{(8f)^{8f}}.
\end{equation*}
With large $n$, this probability is high. That is, for any $\beta >0$, there exists $n^*$, so that for all $n\geq n^*$,

$$\frac{e^{-n\gamma}(en\gamma)^{8f}}{(8f)^{8f}} \leq \beta.$$ 

The threshold $n^*$ is calculated by solving the previous inequality. At this point we obtain another threshold for $n$. Remember that in the end, we will choose the maximum of these thresholds. Hence we can use~\eqref{eq: eps_upp_bound2}.

Next, we make use of the upper bound in~\eqref{eq: hoeffding2}. We set $\exp(-2\epsilon^2 D) = \frac{c}{3}$ and solve it for $D$. Again, remember that we will use the upper bound on $\epsilon$,~\eqref{eq: eps_upp_bound2}:

\begin{equation*}
    \exp(-2\epsilon^2 D) = \frac{c}{3} \geq \exp(-2(\frac{\Tilde{\delta}(f)}{4f})^2 D).
\end{equation*}
We end up with a lower bound on $D$, which we denote by $D^*$,
\begin{equation}\label{lower_bound2}
    D \geq \frac{f^2}{\Tilde{\delta}^2(f)} 8\log(\frac{3}{c}) =: D^*.
\end{equation}

Then, for any $D\geq D^*$, we have $P[X_2 \leq a_2]<\frac{c}{3}$ and hence,
\begin{align}\label{eq: eq: G<c/3_case2}
    \begin{split}
        G_f(D,V) &= P[X_1 > a_1]+\frac{1}{2}P[X_1 = a_1]-P[X_2 > a_2]-\frac{1}{2}P[X_2 = a_2] \\
        &< 1 - P[X_2 > a_2]\\
        &< 1- (1-\frac{c}{3}) = \frac{c}{3},
    \end{split}
\end{align}
where the last inequality precisely follows from Hoeffding's inequality.
As a last step, it remains to show that the probability that $D\geq D^*$ is high and hence~\eqref{eq: eq: G<c/3_case2} holds. We can use again the inequality of~\cite{Mitzenmacher} for bounding a Poisson random variable, namely, for $D^* < n\gamma$, 
 \begin{equation*}
     P[D\leq D^*] \leq \frac{e^{-n\gamma}(en\gamma)^{D^*}}{(D^*)^{D^*}}.
 \end{equation*} 
 
 Again, with $n$ large enough or by taking
 \begin{equation*}
     \Tilde{\delta}(f) > \frac{2f}{\sqrt{\frac{n\gamma}{2\log(3/c)}}-1},
 \end{equation*}
 we can make ensure that $D^* < n\gamma$ and hence 
 \begin{equation*}
     P[D\geq D^*] = 1-P[D\leq D^*] > 1-\frac{e^{-n\gamma}(en\gamma)^{D^*}}{(D^*)^{D^*}}.
 \end{equation*}
 
 We want to lower-bound the latter by $q$, where $q$ is close to $1$. By the property of the Lambert function, we obtain two thresholds $n_1, n_2$, one for applying $W_0$ and the other for applying $W_{-1}$,  
 \begin{equation*}
         n_{1,2} = -\frac{D^*}{\gamma} W_{0,-1}(-\frac{(1-q)^{1/D^*}}{e}).
 \end{equation*}
We choose the higher threshold.

\item Let $V = f + \sigma(f)$. This part is treated as the previous part, where $V=f+\Tilde{\delta}(f)$. Then, the probability $p$ that a vote is delegated to the well-behaving voters is
    \begin{equation*}
        p = \frac{V}{V+f} = \frac{1}{2} + \frac{\sigma(f)}{2(2f+\sigma(f))}
    \end{equation*}
    and $a_2$, as given in~\eqref{eq: a12}, is
    \begin{equation*}
        a_2 = \frac{f+D+1-V}{2} = \frac{D+1+\sigma(f)}{2}.
    \end{equation*} 
    We want to upper bound the following: $P[X_2\leq a_2] < c/3$. We use Hoeffdings's inequality for some real-valued $\epsilon$,
    \begin{equation} \label{eq: hoeffding3}
        P[X \leqslant D(p-\epsilon)] \leqslant \exp(-2 \epsilon^2 D).
    \end{equation} 
    To use~\eqref{eq: hoeffding3}, we need to find $\epsilon$. By setting $D(p - \epsilon) = a_2$, we can solve $\epsilon$: 
    \begin{equation}\label{eq: eps3}
        \epsilon = p - \frac{a_2}{D} = \frac{\sigma(f)}{4f + 2 \sigma(f)}- \frac{\sigma(f)+1}{2D}.
    \end{equation}
Again, for $D > 8f$, we can upper-bound $\epsilon$ as we did it before,
\begin{equation}\label{eq: eps_upp_bound3}
    \epsilon \leq \frac{\sigma(f)}{4f}.
\end{equation}
By the Poisson random variable concentration bound, for $8f < n \gamma$, we can lower-bound the probability that $D > 8f$,
\begin{equation*}
    P[D > 8f]\geq 1- 
    \frac{e^{-n\gamma}(en\gamma)^{8f}}{(8f)^{8f}}.
\end{equation*}
For large $n$, this probability is high. That is, for any $\beta >0$, there exists $n^*$ such that for all $n\geq n^*$:

$$\frac{e^{-n\gamma}(en\gamma)^{8f}}{(8f)^{8f}} \leq \beta.$$ 

The threshold $n^*$ is calculated by solving the above inequality. We obtain one more threshold for $n$. We will finally choose the maximum of all thresholds. Hence we can use~\eqref{eq: eps_upp_bound3}.

Next, we make use of the upper bound in Hoeffding's inequality,~\eqref{eq: hoeffding3}. We set $\exp(-2\epsilon^2 D) = \frac{c}{3}$ and solve it for $D$. We again use the upper bound on $\epsilon$,~\eqref{eq: eps_upp_bound3}:
\begin{equation*}
    \exp(-2\epsilon^2 D) = \frac{c}{3} \geq \exp(-2(\frac{\sigma(f)}{4f})^2 D).
\end{equation*}

We end up with a lower bound on $D$, which we again call $D^*$, 
\begin{equation}\label{lower_2}
    D \geq 8 \frac{f^2}{\sigma^2(f)} \log(\frac{3}{c}) =: D^*.
\end{equation}

Then, for any $D\geq D^*$ we have $P[X_2 \leq a_2]<\frac{c}{3}$ and hence, 
\begin{align}\label{eq: eq: G<c/3_case2.2}
    \begin{split}
        G_f(D,V) &= P[X_1 > a_1]+\frac{1}{2}P[X_1 = a_1]-P[X_2 > a_2]-\frac{1}{2}P[X_2 = a_2] \\
        &< 1 - P[X_2 > a_2]\\
        &< 1- (1-\frac{c}{3}) = \frac{c}{3},
    \end{split}
\end{align}
where the last inequality precisely follows from Hoeffding's inequality.
As the last step, it remains to show that the probability for $D\geq D^*$ is high and hence that~\eqref{eq: eq: G<c/3_case2.2} holds. For this, we can again use the concentration bound on the Poisson random variable, in particular, for $D^* < n\gamma$, 
 \begin{equation*}
     P[D\leq D^*] \leq \frac{e^{-n\gamma}(en\gamma)^{D^*}}{(D^*)^{D^*}}.
 \end{equation*} 
Again, with $n$ large enough or by taking 
\begin{equation*}
    \sigma(f) > \frac{2f}{\sqrt{\frac{n\gamma}{2\log(3/c)}}-1},
\end{equation*}
we can ensure that $D^* < n\gamma$ and hence,
\begin{equation*}
     P[D\geq D^*] = 1-P[D\leq D^*] > 1-\frac{e^{-n\gamma}(en\gamma)^{D^*}}{(D^*)^{D^*}}.
\end{equation*}

We want the latter to be at least $q$, where $q$ is close to $1$.

Therefore, we obtain two thresholds $n_1, n_2$, one by applying $W_0$ and the other by applying $W_{-1}$, 
 \begin{equation*}
         n_{1,2} = -\frac{D^*}{\gamma} W_{0,-1}(-\frac{(1-q)^{1/D^*}}{e}).
 \end{equation*}
We choose the higher threshold.

\end{itemize}

\textbf{Case 3. } Let $V \in (f-\delta(f),f+\Tilde{\delta}(f))$. We can bound the probability that $V$ is in this interval by 
\begin{equation}\label{eq: case3}
    P[V \in (f-\delta(f),f+\Tilde{\delta}(f))] \leq 2\delta(f) \cdot \frac{(f-\delta(f))^{f-\delta(f)}}{e^{f-\delta(f)} (f-\delta(f))!}.
\end{equation} 
We want to obtain the following upper bound:
\begin{equation*}
    P[V \in (f-\delta(f),f+\Tilde{\delta}(f))]<\frac{c}{3}.
\end{equation*}
To obtain this upper bound, we first recall Stirling's inequality. For any $d$, the following holds:
\begin{equation*}
    \sqrt{2 \pi} d^{d+1/2} e^{-d+1/(12d+1)} < d!.
\end{equation*}
Let us rewrite~\eqref{eq: case3}, using $d:= f-\delta(f)$, and apply Stirling's inequality:
\begin{equation*}
     P[V \in (f-\delta(f),f+\Tilde{\delta}(f))] \leq 2\delta(f) \frac{d^d}{e^d d!} < \frac{2\delta(f)}{\sqrt{2 \pi d} e^{1/(12d+1)}}.
\end{equation*}
Next, we set the RHS equal to $c/3$ and obtain
\begin{equation*}
    \delta^2(f) = \frac{\pi c^2}{18} (f-\delta(f)) e^{2/(12(f-\delta(f))+1)}.
\end{equation*}
We can upper-bound $\delta$ by the following:
\begin{equation*}
    \delta(f) \leq c e \sqrt{ \frac{\pi}{18}} \sqrt{f}.
\end{equation*}

The latter can be rewritten as $$f\geq \frac{18}{e^2 \pi}\frac{\delta^2(f)}{c^2}.$$ $\delta(f)$ should be at least $1$. This is a requirement to have $V$ strictly smaller than $f$. By taking the maximum of lower bounds on $D$,~\eqref{lower_bound1},~\eqref{lower_bound2} and~\eqref{lower_2}, we obtain a lower bound for the value of $n$, too. 

Altogether, we have shown that in each case, $G_f(D,V)< \frac{c}{3}$.
This ends the proof of the theorem.  
\end{proof}

\begin{proof}[Proof of Proposition \ref{lem:f1gamm1}]
The right-hand side of equation~\eqref{eq: equicond} for $f=1$ reads
\begin{align*}
\xi_{n,1}(\gamma)&=\sum_{D=0}^\infty \sum_{V=0}^\infty \frac{(n\gamma)^D}{e^{n\gamma}D!} \frac{(n(1-\gamma))^V}{e^{n(1-\gamma)}V!} \times \\& \quad \times\bigg[ \sum_{h=0}^D \binom{D}{h} \left(\frac{V+1}{V+2}\right)^h \left(\frac{1}{V+2}\right)^{D-h} g(V+1+h, 1+D-h) \\
&\quad - \sum_{h=0}^{D+1} \binom{D+1}{h} \left(\frac{V}{V+1}\right)^h \left(\frac{1}{V+1}\right)^{D+1-h} g(V+h, D+2-h) \bigg].
\end{align*}
For $\gamma = 1$, only the terms remain where $V=0$,
\begin{align*}
\xi_{n,1}(1)=\sum_{D=0}^\infty \frac{n^D}{e^{n}D!}\bigg[ \sum_{h=0}^D \binom{D}{h} \left(\frac{1}{2}\right)^D g(1+h, 1+D-h) \bigg].
\end{align*}
Note that $g(1+h, 1+D-h) = \frac{1}{2}$ if $2h = D$ and
$g(1+h, 1+D-h) = 1$ if $2h > D$.
Hence, we have 
\begin{align}\label{eq: lem1}
        \xi_{n,1}(1) 
        &=\sum_{D=0}^\infty \frac{n^D}{e^{n}D!}\bigg[ \binom{D}{h} \left(\frac{1}{2}\right)^D \frac{1}{2} \mathbbm{1}_{\{2h = D\}} + \left(\frac{1}{2}\right)^D \sum_{h=\lceil \frac{D+1}{2} \rceil}^D \binom{D}{h}  \bigg].
\end{align}
We use the following property about the sum of binomial coefficients:
\begin{equation}\label{eq: binomsum}
    \sum_{h=0}^D \binom{D}{h} = 2^D.
\end{equation}
Recall the symmetry property of binomial coefficients for non-negative $D$ and $h$:
\begin{equation}\label{eq: binomsym}
    \binom{D}{h} = \binom{D}{D-h}.
\end{equation}
With the two properties~\eqref{eq: binomsum} and~\eqref{eq: binomsym} we end the proof.

In equation~\eqref{eq: lem1}, the first term in brackets vanishes for odd $D$. This means that for odd $D$, the entire expression in brackets is the following:
\begin{align*}
    \left(\frac{1}{2}\right)^D \sum_{h=\lceil \frac{D+1}{2} \rceil}^D \binom{D}{h} = \left(\frac{1}{2}\right)^D \frac{2^D}{2} = \frac{1}{2},
\end{align*}
where the first equality follows from~\eqref{eq: binomsum} and~\eqref{eq: binomsym}, as we only sum over exactly half of the binomial coefficients.

If $D$ is even, the expression in brackets in equation~\eqref{eq: lem1} is
\begin{align*}
    &  \binom{D}{\frac{D}{2}} \left(\frac{1}{2}\right)^D \frac{1}{2} + \left(\frac{1}{2}\right)^D \sum_{h=\lceil \frac{D+1}{2} \rceil}^D \binom{D}{h} \\ 
    &= \left(\frac{1}{2}\right)^D \left( \binom{D}{\frac{D}{2}} \frac{1}{2} + \sum_{h=\lceil \frac{D+1}{2} \rceil}^D \binom{D}{h} \right)\\
    &= \left(\frac{1}{2}\right)^D  \frac{2^D}{2} = \frac{1}{2},
\end{align*}
where again, the second equality follows from~\eqref{eq: binomsum} and~\eqref{eq: binomsym}.

It follows that
\begin{equation*}
    \xi_{n,1}(1) = \frac{1}{2} \sum_{D=0}^\infty \frac{n^D}{e^{n}D!} = \frac{1}{2}.
\end{equation*}

The right-hand side of equation~\eqref{eq: equicond} for $f=1$ reads
\begin{align*}
\xi_{n,1}(\gamma)&=\sum_{D=0}^\infty \sum_{V=0}^\infty \frac{(n\gamma)^D}{e^{n\gamma}D!} \frac{(n(1-\gamma))^V}{e^{n(1-\gamma)}V!} \times \\ & \quad \times \bigg[ \sum_{h=0}^D \binom{D}{h} \left(\frac{V+1}{V+2}\right)^h \left(\frac{1}{V+2}\right)^{D-h} g(V+1+h, 1+D-h) \\
&\quad - \sum_{h=0}^{D+1} \binom{D+1}{h} \left(\frac{V}{V+1}\right)^h \left(\frac{1}{V+1}\right)^{D+1-h} g(V+h, D+2-h) \bigg].
\end{align*}
For $\gamma = 0$, only the terms remain where $D=0$,
\begin{align*}
\xi_{n,1}(0)&=\sum_{V=0}^\infty \frac{n^V}{e^{n}V!}\bigg[ g(V+1,1) - \sum_{h=0}^{1} \binom{1}{h} \left(\frac{V}{V+1}\right)^h \left(\frac{1}{V+1}\right)^{1-h} g(V+h, 2-h) \bigg] \\ 
&= \sum_{V=0}^\infty \frac{n^V}{e^{n}V!}\bigg[ g(V+1,1) - \frac{1}{V+1} g(V,2) - \frac{V}{V+1} g(V+1, 1) \bigg].
\end{align*}
From the equation above, we see immediately that the term in brackets is $0$ for $V\geq 3$, as all $g(\cdot, \cdot)$-terms are $1$.
Therefore, we only need to consider $\xi_{n,1}(0)$ for $V=0,1,2$. It follows that
\begin{equation*}
    \xi_{n,1}(0) = \frac{1}{2e^n} + \frac{n}{2e^n} + \frac{n^2}{12e^n} = \frac{1}{e^n} (\frac{1}{2}+\frac{n}{2}+\frac{n^2}{12}).
\end{equation*}
Since $\xi_{n,1}(\gamma)$ is a continuous function, equation~\eqref{eq: equicond} has a solution for any $c\in [\frac{1}{e^n} (\frac{1}{2}+\frac{n}{2}+\frac{n^2}{12}),\frac{1}{2}]$. 
\end{proof}

\begin{proof}[Proof of Corollary \ref{prop: base}]
The derivative w.r.t. $\alpha$ of the RHS of~\eqref{eq: equicond_base} is given by
\begin{equation*}
    \frac{n^f}{2f!}\frac{f \alpha^{f-1}-n \alpha^f}{e^{n \alpha}} + \frac{n^{f-1}}{2(f-1)!}\frac{(f-1) \alpha^{f-2}-n \alpha^{f-1}}{e^{n \alpha}}. 
\end{equation*}
Setting this equation equal to $0$, we obtain
\begin{align*}
    0 &= n^f(f \alpha^{f-1} - n \alpha^f) + n^{f-1}f((f-1) \alpha^{f-2} - n \alpha^{f-1})\\
    &= \alpha^{f-2}(f(f-1)-n^2 \alpha^2).
\end{align*}
The only positive solution is $\alpha^* = \frac{\sqrt{f(f-1)}}{n}$. We can easily verify that the RHS of~\eqref{eq: equicond_base} has a maximum at $\alpha^*$ by inserting $\alpha^*$ into the second derivative of the RHS of~\eqref{eq: equicond_base}. As a last step, we have to insert $\alpha^*$ into the RHS of~\eqref{eq: equicond_base} to obtain the right endpoint of the interval. As for $f\geq 2$ and $\alpha = 0$, the RHS of~\eqref{eq: equicond_base} is $0$, the left endpoint of the interval is $0$.

For $f=1$, we simply insert $\alpha = 0$ and $\alpha = 1$ into the RHS of~\eqref{eq: equicond_base} to obtain both endpoints of the interval.
\end{proof}

\end{document}